\documentclass[a4paper,fleqn,usenatbib,aas_macros]{mnras}

\usepackage{mathptmx}

\usepackage[T1]{fontenc}
\usepackage{ae,aecompl}
\usepackage{tikz}
\usetikzlibrary{positioning}

\usepackage{graphicx}	
\usepackage{amsmath}	
\usepackage{amssymb}	
\usepackage{bm}         
\usepackage{color}
\newcommand{\bc}{}

\title[External photoevaporation]{A slim disc approach to external photoevaporation of discs}

\author[Owen, J. E. \& Altaf, N.]{
James E. Owen\thanks{E-mail: james.owen@imperial.ac.uk} and Noumahn Altaf
\\
Astrophysics Group, Department of Physics, Imperial College London, Prince Consort Rd, London, SW7 2AZ, UK
}

\pubyear{2015}

\begin{document}
\label{firstpage}
\pagerange{\pageref{firstpage}--\pageref{lastpage}}
\maketitle

\begin{abstract}
The photoevaporation of protoplanetary discs by nearby massive stars present in their birth cluster plays a vital role in their evolution. Previous modelling assumes that the disc behaves like a classical Keplerian accretion disc out to a radius where the photoevaporative outflow is launched. There is then an abrupt change in the angular velocity profile, and the outflow is modelled by forcing the fluid parcels to conserve their specific angular momenta. Instead, we model externally photoevaporating discs using the slim disc formalism. The slim disc approach self consistently includes the advection of radial and angular momentum as well as angular momentum redistribution by internal viscous torques. Our resulting models produce a smooth transition from a rotationally supported Keplerian disc to a photoevaporative driven outflow, where this transition typically occurs over $\sim$4-5 scale heights. The penetration of UV photons predominately sets the radius of the transition and the viscosity's strength plays a minor role. By studying the entrainment of dust particles in the outflow, we find a rapid change in the dust size and surface density distribution in the transition region due to the steep gas density gradients present. This rapid change in the dust properties leaves a potentially observable signature in the continuum spectral index of the disc at mm wavelengths. Using the slim disc formalism in future evolutionary calculations will reveal how both the gas and dust evolves in their outer regions and the observable imprints of the external photoevaporation process.

\end{abstract}



\section{Introduction}
The majority of stars, and hence the planets that form around them are born in star clusters. Therefore, the {\it environment} in which stars form can influence the evolution of their discs and perhaps any subsequent planetary system. In particular, two consequences for protoplanetary discs evolution in young stellar clusters have been identified: tidal truncation and photoevaporative mass-loss. 

If the young cluster is sufficiently dense, tidal interactions with other stars due to close-encounters can truncate and remove material from discs \citep[e.g.][]{Clarke1993,Pfalzner2005,Rosotti2014,Winter2018}. Additionally, heating due to UV radiation from nearby massive stars can drive powerful photoevaporative outflows that remove material from the disc \citep[e.g.][]{Johnstone1998,Adams2004,Facchini2016,Haworth2018}. Ultimately, both these processes reduce the disc's size (shortening its viscous timescale) and remove mass, hastening the discs eventual dispersal \citep[e.g.][]{Parker2020}. While both processes have been observed to be occurring in real star forming regions \citep[e.g.][]{Bally2000,Facchini2016b}, the structure of young clusters implies that, in general, external photoevaporation plays the more important role in promoting disc dispersal \citep[e.g.][]{Winter2020}. 

While external photoevaporation was first identified and studied with respect to extreme UV environments such as the Orion Nebula Cluster \citep[e.g.][]{Odell1993,Johnstone1998,Scally2001,Adams2004,Clarke2007,Anderson2013}, work has recently extended the photoevaporation models to less extreme UV environments indicating it still plays and important role \citep[e.g.][]{Kim2016,Facchini2016,Haworth2017,Haworth2018,Concha2019,HO2020,Parke2021}, particularly for discs around very low-mass stars \citep[e.g.][]{Haworth2018b}.  

Models for the photoevaporative outflow itself and their impact on the evolution are distinct. Specifically, (normally 1D) hydrodynamic simulations are used to pre-calculate mass-loss rates as a function of disc and environmental parameters. This mass-loss is then included as a sink term in the evolution of a disc's surface density \citep[e.g.][]{Clarke2007,Anderson2013,Selleck2020} at the disc's outer edge. Physically, this implementation implies the disc transitions instantly to the photoevaporative outflow and there is zero viscous torque across this interface. A zero torque outer boundary will artificially steepen the decline in surface density in classic disc models. Further, the photoevaporative outflows neglect the treatment of viscous angular momentum transport and assume the fluid parcels conserve their specific angular momentum. While such an implementation is likely to be representative of reality in extreme UV environments where the high EUV fluxes results in a sharp ionization front with an extreme temperature jump at the disc's outer edge. However, it's unclear whether this parameterisation is true for more moderate environments, where FUV heating results in a more gradual change of  the temperature of ionization structure \citep[e.g.][]{Facchini2016,Haworth2018}. While at the broad-brush level such an implementation is likely to be accurate for the evolution of the gas disc, the relative drift and entrainment of dust is sensitive to the details of the gas pressure gradient and angular velocity distribution \citep[e.g.][]{takeuchi02}. Furthermore, detailed comparison of the kinematics of the outer disc edges by {\it ALMA} \citep[e.g.][]{Haworth2017,HO2020} to photoevaporative models are likely to be sensitive to the disc/outflow interface. 

Previous models of external photoevaporation treat the disc and outflow differently because the assumptions made by the disc model (Keplerian, neglecting pressure gradients) and the outflow model (fluid particles conserve specific angular momentum) are incompatible. In this work, we make use of ``slim'' disc models \citep{Abramowicz1988}, a class of accretion disc models that include the necessary physics to model the transition from a Keplerian disc to a photoevaporative outflow smoothly. Section~2 includes an implementation of the slim disc model for external photoevaporating discs. Section~3 includes an application to globally isothermal models. Section~4 considers non-isothermal models more representative of real photoevaporating systems, and Section~5 considers how dust particles evolve in these discs/outflows. We discuss our results in Section~6 and summarise in Section 7.

\section{Overview: the slim disc model}

\begin{figure*}
    \centering
    \includegraphics[width=\textwidth,trim={0 19cm 0 0},clip]{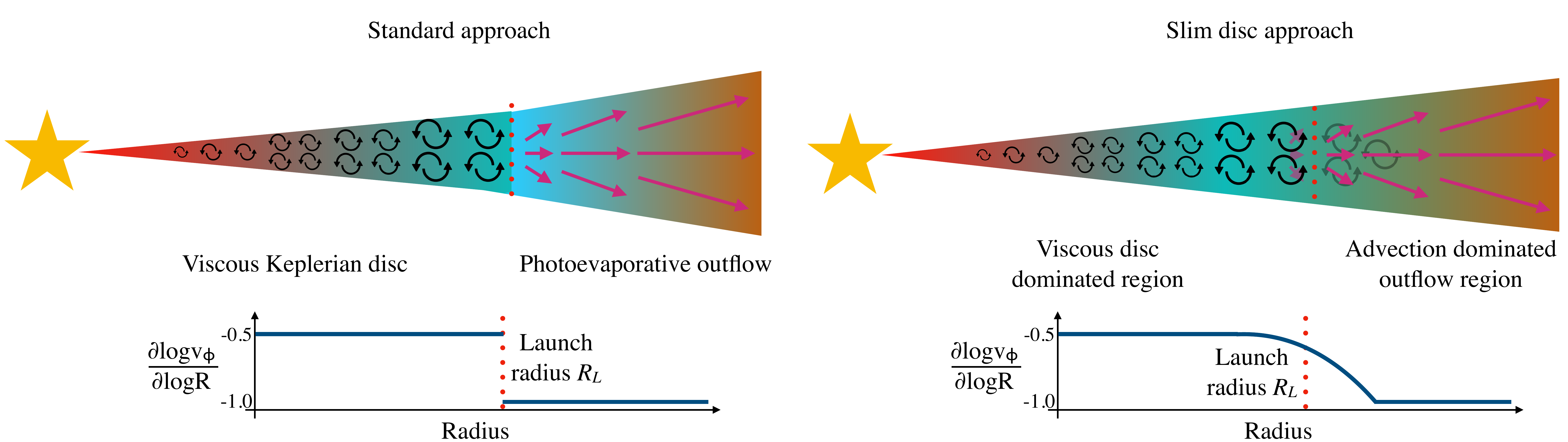}
    \caption{Schematic diagram showing the differences between the current, standard approach, where there is an abrupt change in the angular velocity profile and modelling philosophy at the launch radius (left) and our slim disc approach where there is a smooth transition. In our slim disc approach, advection of momentum can act inside the launch radius and viscosity can operate outside. In the standard approach, viscosity exclusively acts to transport material inside the launch radius and advection of momentum exclusively operates outside the launch radius. The bottom panels indicate the evolution of the angular velocity profile as a function of radius evolving from a Keplerian profile ($\partial\log v_\phi/\partial \log R = -1/2$), to a profile where the specific angular momentum is radially constant  ($\partial\log v_\phi/\partial \log R = -1$). We define the launch radius ($R_L$) to be the radius at which the fluid's specific angular momentum at a large distance into the outflow is given by $\sqrt{GM_*R_L}$.}
    \label{fig:diagram}
\end{figure*}

Protoplanetary discs undergoing external photoevaporation behave like ``classical'' thin accretion discs\footnote{Provided angular momentum transport is provided by internal, local torques.} close to their stars, whereas at large radii in the photoevaporative flow, redistribution of angular momentum is unimportant and fluid parcels can conserve their angular momentum. In the standard approach to thin accretion discs \citep[e.g.][]{Pringle1972,LyndenBellPringle1974,Pringle1981}, the assumption that the disc's scale height ($H\approx c_s/\Omega_K$, where $c_s$ is the isothermal sound speed, and $\Omega_K$ is the Keplerian angular velocity), is small compared to the radial location implies that both the radial pressure support and radial advection of linear and angular momentum are negligible. Under this set of approximations, the angular velocity profile ($\Omega$) of the disc is described by the Keplerian profile ($\Omega = \Omega_K = \sqrt{GM_*/R^3}$, with $G$ the gravitational constant, $M_*$ the star's mass and $R$ the radial distance).  However, in the photoevaporative outflow, which by its physical origin means that radial pressure support dominates,  the advection of linear and angular momentum dominate over viscous processes. In previous models of the photoevaporative outflow, redistribution of angular momentum by an internal torque is neglected and the angular velocity profile is given by $\Omega = h/R^2$, where $h$ is the specific angular momentum carried by the fluid parcels. Clearly, the photoevaporative outflow cannot be accommodated within the approximations made by the thin disc model, nor can the redistribution of angular momentum by internal torques be incorporated into the current approach to external photoevaporation. 

However, the ``slim'' disc formalism \citep{Abramowicz1988} provides an approach to capture both the accretion disc component and the photoevaporative flow along with modelling the transition between the two. The differences between the current, ``standard'' approach, and our slim disc approach are schematically shown in Figure~\ref{fig:diagram}.

At this initial stage we only consider steady-state solutions (but discuss their limitations in Section~\ref{sec:limitations}), where the mass-flow rate, $\dot{M}$, is radially constant and the disc is vertically isothermal. In the slim disc approach, force balance in the vertical direction still yields\footnote{Note that assumption of force balance in the vertical direction likely breaks down close to the sonic point; however, as we discuss in Section~\ref{sec:limitations} this is not near the transition from disc dominated to outflow dominated and does not effect our modelling at this point.}:
\begin{equation}
    H = \frac{c_s}{\Omega_K} \label{eqn:H_def}
\end{equation}
and force balance in the azimuthal direction still yields (assuming a Navier-Stokes like viscous stress):
\begin{equation}
    \rho u_R \frac{{\rm d}h}{{\rm d}R} = \frac{1}{R}\frac{\partial}{\partial R} \left(R^3 \mu \frac{{\rm d}\Omega}{{\rm d}R}\right) \label{eqn:force_az}
\end{equation}
where $\rho$ is the mass-density, $u_R$ is the radial velocity and $\mu$ is the effective dynamical viscosity that parameterises the internal (turbulent) torques. However, force balance in the radial direction now becomes:
\begin{equation}
    u_R\frac{\partial u_R}{\partial R} = -\frac{1}{\rho}\frac{\partial P}{\partial R} + \left(\Omega^2 - \Omega_K^2\right)R \label{eqn:force_r}
\end{equation}
while Equations~\ref{eqn:H_def} and \ref{eqn:force_az} have identical forms to the standard thin-accretion disc model, the fact that $\Omega \ne \Omega_K$ from radial force balance means the Equation~\ref{eqn:force_az} evolves differently. Vertical integration of Equation~\ref{eqn:force_az} yields:
\begin{equation}
    \Sigma v_R \frac{{\rm d}h}{{\rm d}R} = \frac{1}{R}\frac{\partial}{\partial R} \left(\nu \Sigma R^3 \frac{{\rm d}\Omega}{{\rm d}R}\right) \label{eqn:vert_int_az}
\end{equation}
where $\Sigma$ is the gas surface density, $v_R$ is the vertically averaged velocity, defined such that the mass flow rate is given by $\dot{M}=2\pi R\Sigma v_R$ and $\nu$ is a density-weighted average kinematic viscosity. Vertical averaging of Equation~\ref{eqn:force_r} is not a trivial procedure, as in a real disc both the convective radial derivative and pressure gradient can strongly vary with height, although as argued by \citet{Abramowicz1988} the results only weakly depend on how this vertical averaging is done. Since most of the disc's mass is contained near the mid-plane we choose to evaluate the pressure term at the mid-plane and obtain:
\begin{equation}
    v_R\frac{\partial v_R}{\partial R} = -\frac{H}{\Sigma}\frac{\partial \left(\Sigma c_s^2 /H\right)}{\partial R} + \left(\Omega^2 - \Omega_K^2\right)R \label{eqn:vert_int_r}
\end{equation}
As will become clear in Section~\ref{sec:isothermal} this choice maximises the impact of the photoevaporative outflow.

\subsection{Numerical scheme}

In later sections we search for numerical solutions to our disc model. While previous authors have solved this problem with a relaxation code \citep[e.g.][]{Popham1991,Krticka2011}, this quickly becomes numerically challenging as increasing complex physical processes are included (such as ray-tracing and dust physics). Instead, even though we are after steady solutions, we follow the approach of \citet{Kurfurst2014} and insert time-dependent terms into Equations~\ref{eqn:vert_int_az} and \ref{eqn:vert_int_r}, along with the continuity equation:
\begin{equation}
    \frac{\partial \Sigma}{\partial t} +\frac{1}{R}\frac{\partial}{\partial R}\left(Rv_R\Sigma\right) = 0
\end{equation}
and solve them using an explicit, time-dependent scheme. We use a first-order in time, second-order in space {\sc zeus}-like algorithm \citep{Stone1992}, where we use van-Leer slope limiters, a von Neumann \& Richtmyer artificial viscosity and a CFL condition of 0.35. The turbulent viscosity term is included in an extra sub-step before the transport-step through operator splitting. We then evolve the time-dependent model to steady-state to obtain our steady solutions.   
\section{Insights from isothermal models}\label{sec:isothermal}
We can use isothermal models to gain critical insights into the transition from a viscously dominated disc to an outflow. If the sound-speed is globally constant, Equation~\ref{eqn:vert_int_r} combined with the continuity equation becomes:
\begin{equation}
    \left(\mathcal{M}-\mathcal{M}^{-1}\right) \frac{\partial \mathcal{M}}{\partial\log R} = 1 + \frac{\partial \log H}{\partial \log R} + \left(\frac{v_K}{c_s}\right)^2\left(\frac{v_\phi^2}{v_K^2}-1\right)
\end{equation}
where $\mathcal{M}=v_R/c_s$ is the radial Mach number of the flow. Since $H$ generally increasing with distance for real discs we note the use of the mid-plane pressure gradient in our vertically averaged expressions (which introduces the $\partial\log H/\partial \log R$ term) maximises the power of the pressure driven outflow.  Using $\partial\log H/\partial \log R = 3/2$ for our isothermal choice, we get:
\begin{equation}
 \left(\mathcal{M}-\mathcal{M}^{-1}\right) \frac{\partial \mathcal{M}}{\partial\log R} = \frac{5}{2} + \left(\frac{v_K}{c_s}\right)^2\left(\frac{v_\phi^2}{v_K^2}-1\right) \label{eqn:isothermal_mom}
\end{equation}
As expected Equation~\ref{eqn:isothermal_mom} contains a critical point at the sonic point. In the absence of rotation we obtain the sonic point location at:
\begin{equation}
    R_s = \frac{2GM_*}{5c_s^2} \equiv R_{s0}
\end{equation}
which is analogous to the Parker wind solution \citep{Parker1958}, with a smaller coefficient due to the super-spherical velocity divergence we have assumed in our disc model. 

\subsection{Outflow dominated region}

Now in the outflow dominated region, where viscosity is unimportant, the gas-parcels conserve specific angular momentum and we can write $v_\phi=h_w/R$, where $h_w$ is the specific angular momentum carried away by the outflow. Identifying this with the ``launch'' radius ($R_L$) of the outflow we can approximately label the radius at which the outflow begins to dominate over viscous transport by defining $R_L \equiv h_w^2/GM_*$. Under the assumption that $h_w$ is constant in the outflow dominated region we can again solve Equation~\ref{eqn:isothermal_mom} for the sonic point location to find:
\begin{equation}
    R_s^2 - R_{s0}R_s + \frac{2h_w^2}{5c_s^2} = 0 \label{eqn:sonic_quad}
\end{equation}
This quadratic equation has two solutions; however only the larger also satisfies the condition that $\partial \mathcal{M}/\partial R > 0$, as necessary for the actual sonic point. Thus, we find:
\begin{equation}
    R_s = \frac{R_{s0} + R_{s0}\sqrt{1 -4\left(\frac{R_L}{R_{s,0}}\right)}}{2}\approx R_{s0} - R_L \label{eqn:sonic_sol}
\end{equation}
where we see that introduction of rotation in the outflow makes a slight decreases the distance to the sonic-point \citep[e.g.][]{Facchini2016}.  

The alternative solution to Equation~\ref{eqn:sonic_quad} represents the situation where $\partial \mathcal{M}/\partial R = 0$, which occurs approximately at $R_L$. This is actually simple to understand: at $R_L$ the centrifugal and gravitational forces balance, at radii less than $R_L$ there is a net inward acceleration (as gravity is stronger than the centrifugal force) thus any outflow will decelerate with as it approaches $R_L$, and then can accelerate outside $R_L$ as gravity is now weaker than the centrifugal force. Thus, unsurprisingly $R_L$ approximately represents the slowest position in any outflow. 

With constant specific angular velocity Equation~\ref{eqn:isothermal_mom} can be integrated to find the velocity profile:
\begin{equation}
    v_{R,{\rm wind}} = c_s \sqrt{-W\left\{-\left(\frac{R_s}{R}\right)^5\exp\left[f(R)\right] \right\}}
\end{equation}
where $W$ is the Lambert W function \citep[e.g.][]{Cranmer2004}, and
\begin{equation}
    f(R) = \left(\frac{2GM_*}{c_s^2}\right)\left(\frac{1}{R_s}-\frac{1}{R}\right) -\frac{h_w^2}{c_s^2}\left(\frac{1}{R_s^2}-\frac{1}{R^2}\right) -1
\end{equation}

\subsection{Viscosity dominated region}
In the disc dominated region, we instead assume that the angular velocity remains Keplerian, and model the disc as a decretion disc \citep[e.g.][]{Pringle1991} which drives a steady mass and angular momentum flux outwards. We can obtain a solution for the disc dominated region by direct integration of Equation~\ref{eqn:vert_int_az} up to the launch radius $R_L$, assuming a zero torque boundary condition at $R_L$ (as adopted in the standard approach):
\begin{equation}
    R\Sigma v_R \int_R^{R_L} \frac{{\rm d}h}{{\rm d}R}{\rm d}R = \int_R^{R_L} \frac{\partial}{\partial R} \left(\nu \Sigma R^3 \frac{{\rm d}\Omega}{{\rm d}R}\right) {\rm d}R
\end{equation}
which yields:
\begin{equation}
    v_{R,{\rm visc}} = \frac{3 R \nu \Omega_K}{2\left(h_w-\Omega_KR^2\right)}
\end{equation}
This clearly shows that the velocity rapidly increases and diverges as the gas fluid approaches $R_L$. As discussed above $R_L$ also represents the slowest position in any outflow that conserves specific angular momentum. This means that, as expected $R_L$ represents the natural transition between the viscosity dominated region, where internal torques can drive a sufficiently powerful outflow, to one where a thermally driven outflow takes over. 

\subsection{Transition from disc to outflow}
We can understand the transition from the disc dominated region to the outflow by approximately solving for the launch radius ($R_L$). By matching the viscous dominated solution to the outflow solution, we can solve for how $R_L$ changes with the disc's properties. We do this as follows, we approximately write the viscous driven velocity at $R_L$ as: 
\begin{equation}
    v_R\approx \frac{3R_L\nu \Omega_K}{2h_w}  = \frac{3}{2}\alpha \frac{H}{R_L} c_s \label{eqn:vr_approx1}
\end{equation}
where we have made use of the alpha prescription for kinematic viscosity $\nu=\alpha c_s H$ in the final equality. Since $R_L\ll R_s$, we can approximate $R_s\approx R_{s0}$ and,
\begin{equation}
    f(R)\approx-\left(\frac{2GM_*}{c_s^2R_L}\right)
\end{equation}
since $f(R)$ is large and negative, the argument inside the Lambert W-function is small ($\ll 1$) indicating the appropriate branch is the 0 branch, which has a Taylor expansion $W_0(x)\approx x$ for $|x|\ll 1$. Indicating the outflow velocity due to the for the outflow region can be approximately written as:
\begin{equation}
    v_R \approx \left(\frac{R_{s0}}{R_L}\right)^{5/2}\exp\left(-\frac{5R_{s0}}{2R_L}\right) c_s \label{eqn:vr_approx2}
\end{equation}
Thus for mass-continuity from the viscosity dominated region to the outflow dominated region we can equate Equation~\ref{eqn:vr_approx1} and \ref{eqn:vr_approx2} to find:
\begin{equation}
    R_L \approx -\frac{5 R_{s0}}{12W_{-1}\left[-\frac{5}{12}\left(\sqrt{\frac{9}{10}}\alpha\right)^{1/3}\right]} 
\end{equation}
where $W_{-1}$ is the minus one branch of the Lambert W function. As $\alpha$ is small, we can approximate the Lambert W function as $W_{-1}(x) \approx 0.8 \log (-x)$ \citep{Garaud2007}\footnote{Although \citet{Garaud2007} provides this approximation as $\lim_{x\to\infty}W_0(x)$, the symmetry of the Lambert W function means $\lim_{x\to\infty}W_0(x)$ $\longleftrightarrow\;$ $\lim_{x \to 0^+}W_{-1}(-x)$.}, which yields a slightly more intuitive result:
\begin{equation}
    R_L \approx -\frac{R_{s0}}{\log\left(0.069 \alpha\right)}  \label{eqn:approx-launch}
\end{equation}
Thus, we see that the launch radius only has a weak, logarithmic dependence on the strength of the turbulent viscosity. This can be interpreted as follows: the mass-flux carried by the disc only varies linearly with viscosity; however, the mass-flux carried by the outflow varies approximately exponentially with radius (Equation~\ref{eqn:vr_approx2}), thus for continuity of mass-flux the transition position is going to vary logarithmically with viscosity. This implies that the exact value of the viscosity does play some role in setting the transition from the disc dominated region to the outflow, although due to the logarithmic dependence it will be weak.

\subsection{Numerical results}
For the isothermal case, the outflow properties are independent of the choice of density at the inner boundary. As shown above the model can be completely cast in dimensionless parameters; however, since we are concerned with physical systems we'll work in physical units. Thus, we select a solar-mass star, with an outflow composed of molecular gas at a temperature of 150~K, with a mean-molecular weight of 2.35. The inner boundary is held at 20~AU and the outer boundary is located at 1000~AU. We use a grid with 1000 cells logarithmically spaced between the inner and outer boundary. The viscosity law is taken to be the alpha-parameterisation, with a constant value of $\alpha$. The simulation is initialised with a pure decretion disc solution and left to evolve to steady-state. Steady-state is confirmed by an approximately constant angular momentum flux at the outer boundary over a viscous timescale.

\begin{figure}
    \centering
    \includegraphics[width=\columnwidth]{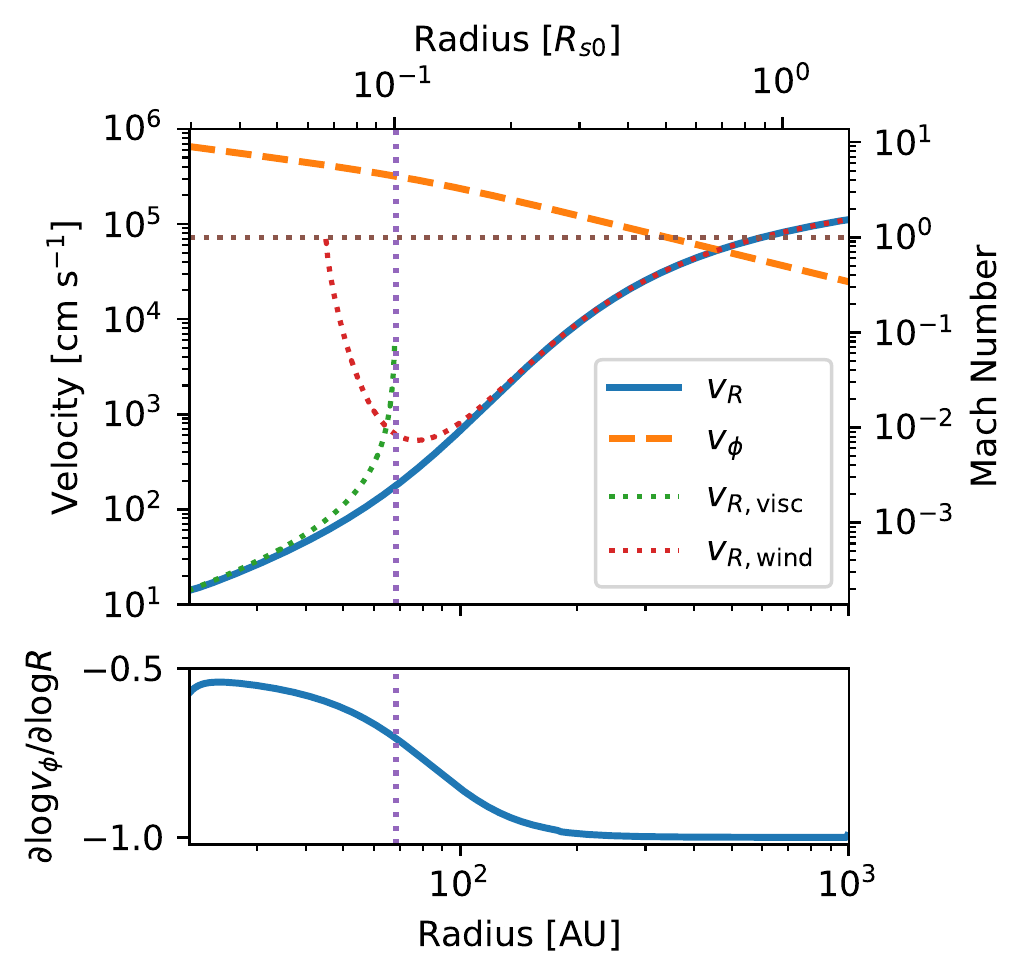}
    \caption{The flow profiles for an isothermal simulation with a viscous $\alpha$ of $10^{-3}$. The vertical dotted line shows the ``launching'' radius of the wind ($R_L$), defined such that the angular momentum at the outer boundary is $h=\sqrt{GMR_L}$. The dotted horizontal line shows the isothermal sound-speed. The top panel shows the velocity profiles in the radial and azimuthal direction, including the approximate solutions in the disc dominated and wind dominated region detailed in Section~\ref{sec:isothermal}. The bottom panel shows how the radial dependence of the azimuthal velocity varies with distance. This plot shows the smooth transition from a Keplerian disc ($\partial \log v_\phi/\partial \log R = -0.5$) to a outflow than conserves specific angular momentum ($\partial \log v_\phi/\partial \log R = -1$).}
    \label{fig:iso_alpha1e-3}
\end{figure}

In Figure~\ref{fig:iso_alpha1e-3} we show the outflow properties for a case where the viscous alpha is $10^{-3}$. As expected we find an outflow that smoothly transitions from a disc dominated decretion outflow, to a pressure driven wind that conserves specific angular momentum. We identify the launching radius, $R_L$, numerically from the simulation by measuring the specific angular velocity at the outer boundary and defining $h_{\rm out}=\sqrt{GMR_L}$. This launching radius naturally sits between the two limiting cases. This point is made clearer in the bottom panel of Figure~\ref{fig:iso_alpha1e-3}, where we show how the radial dependence of the angular velocity varies with radius. At radii $<R_L$ the azimuthal velocity is close to the Keplerian profile $v_{\phi}\propto R^{-1/2}$, where the slightly smaller value comes from the radial pressure support that makes the disc slightly sub-Keplerian. However, at radii $>R_L$ the outflow closely becomes a flow that conserves specific angular momentum,  where $v_{\phi}\propto R^{-1}$. Our simulation indicates that the transition form the disc to outflow is not sharp and occurs over scales of order $R_L$, at least in the isothermal case.  

\begin{figure}
    \centering
    \includegraphics[width=\columnwidth]{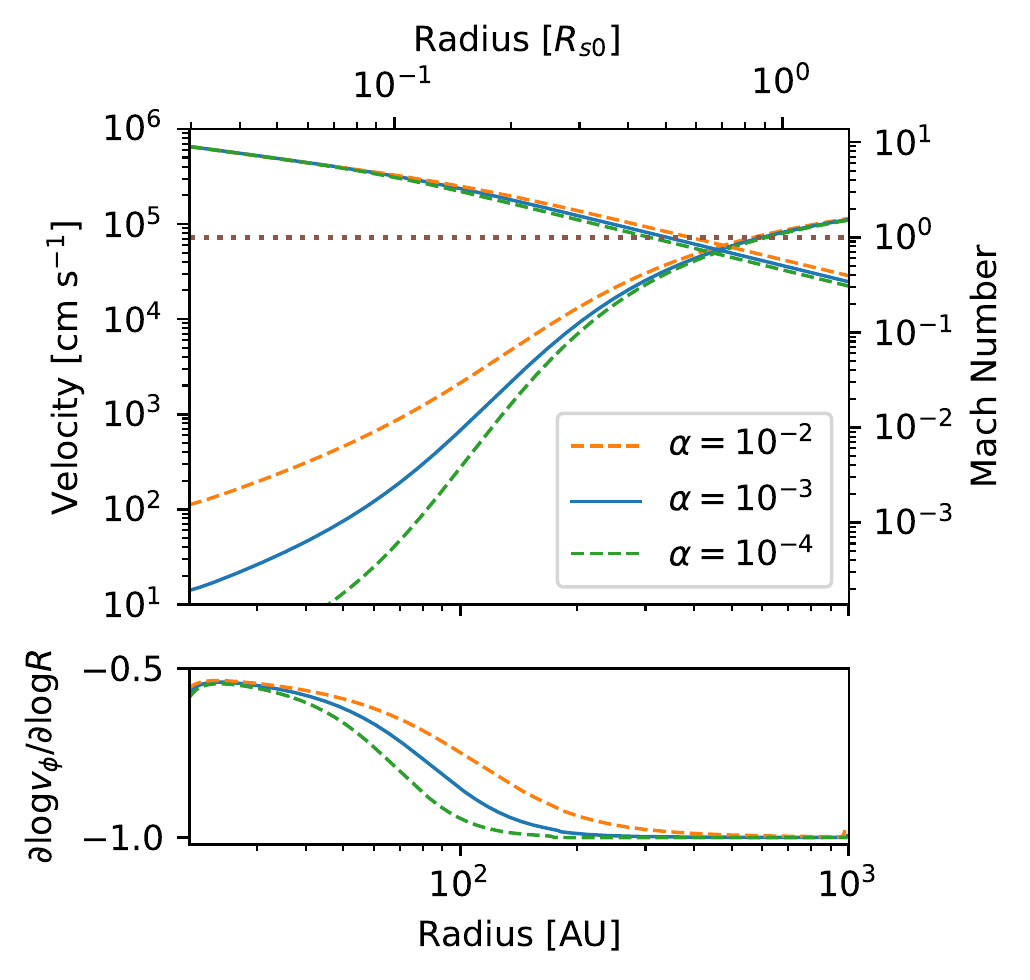}
    \caption{The top panel shows the velocity profiles for the outflowing disc (radially increasing -- radial velocity, radially decreasing -- azimuthal velocity) for different values of the viscous alpha parameter. The bottom panel shows how the radial dependence of the azimuthal velocity varies with distance. The $\alpha=10^{-3}$ is the same flow profile shown in Figure~\ref{fig:iso_alpha1e-3}.}
    \label{fig:varyvisc}
\end{figure}

\begin{figure}
    \centering
    \includegraphics[width=\columnwidth]{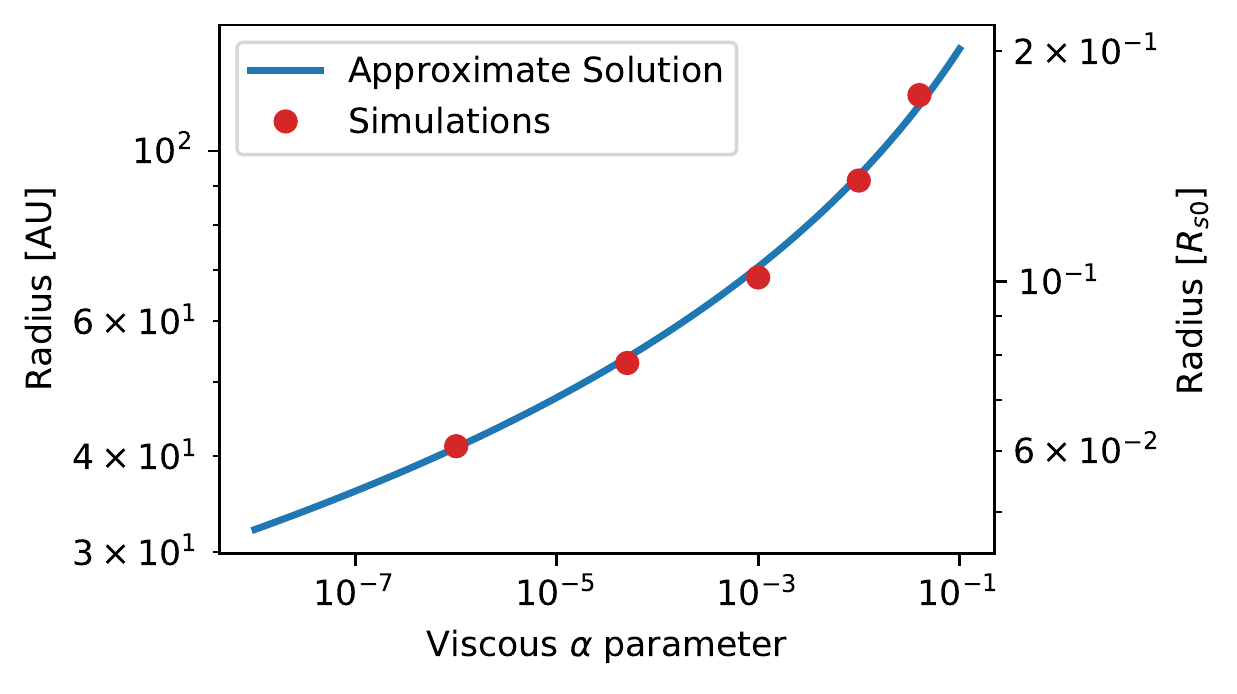}
    \caption{The launch radius ($R_L$) as a function of the viscous alpha parameter for our isothermal case. The line shows the approximate solution from Equation~\ref{eqn:approx-launch}. The points shows the values measured from the simulations assuming that the angular velocity at the outer boundary is given by $\sqrt{GMR_L}$. Clearly, Equation~\ref{eqn:approx-launch} provides a reasonable estimate of the launch radius.}
    \label{fig:Rlvsalpha}
\end{figure}

We can consider how changing the viscosity impacts the outflow solution. In Figure~\ref{fig:varyvisc} we find that by varying the viscosity we vary the transition between the disc dominated and outflow dominated region. However, since the sonic point occurs at approximately $R_{s0}-R_L$ (Equation~\ref{eqn:sonic_sol}), the position of the sonic point only varies slightly (as $R_L\ll R_{s0}$). Finally, the bottom panel of Figure~\ref{fig:varyvisc} indicates that larger values of the viscosity tend to produce smoother transitions. This result is fairly simple to understand since the isothermal outflow solution has $v_R\approx {\rm const}$ close to the launch radius, then additional transport from viscosity can remain competitive over a larger region if the value of the viscosity is larger.   Additionally, we can see how the launch radius varies with viscosity. In Figure~\ref{fig:Rlvsalpha}, we show how the launch radius varies as a function of the viscous alpha parameter. As expected from our discussion in the previous section we only expect a weak, logarithmic dependence. This dependence is indeed what our simulations show, and that our approximate solution from Equation~\ref{eqn:approx-launch} does provide a reasonable representation of the launch radius, with launch radii going to be of order $\sim 0.1 R_{s0}$ for the expected values of the viscosity in protoplanetary discs, indicating that we expect (unless there's a very large jump in temperature due to the external irradiation) most external photoevaporative flows will be launched ``sub-critically'' \citet{Adams2004}. We will make use of these insight when we consider non-isothermal outflows in the next section.

\section{Non-isothermal Models} \label{sec:noniso}
While useful tools for understanding the outflow properties, the isothermal models are unlikely to represent the physical situation for actual protoplanetary discs. In the absence of external irradiation the outer regions of protoplanetary discs are expected to be of order 10~K \citep[e.g.][]{dalessio01}. For molecular hydrogen this gives a value of $R_{s0}$ of $\sim 8,600$~AU, indicating a transition to an outflow dominated region at a radius of $\sim 860$~AU, considerably larger than typical disc sizes \citep[e.g.][]{Ansdell2018}. However, discs subjected to external irradiation, even by weak FUV fields can reach temperatures of $\sim 100$~K and dissociate the molecular hydrogen \citep[e.g.][]{Facchini2016,Haworth2018}, which would give rise to a launching radius of $\sim 50$~AU if the outflow was isothermal (as seen in our isothermal simulations). 

In this work, we do not attempt to model the radiative-transfer, chemistry and hence temperature accurately, as this has been studied extensively \citep[e.g.][]{Adams2004,Facchini2016,Haworth2016,Haworth2018,Haworth2019}. {\bc Rather, we are interested in studying how the transition from a disc-dominated decretion region to a thermally driven outflow happens using our slim disc model. Since the dynamics depend on the isothermal sound-speed (to evaluate the pressure and scale-height), to incorporate the combined influence of the temperature and mean molecular weight change that occurs as a result of UV-irradiation we choose to parameterise the sound-speed of the gas as follows:
\begin{equation}
    c_s = c_{s,{\rm cold}}\left[1-\exp\left(-\tau_{UV}\right)\right] + c_{s,{\rm hot}}\exp\left(-\tau_{UV}\right) \label{eqn:csprofile}
\end{equation}
where $\tau_{UV}$ is the mid-plane optical depth to UV photons and $c_{s,{\rm cold}}$ and $c_{s,{\rm hot}}$ represent the gas' isothermal sound speed in the disc dominated and outflow dominated regions respectively. This simple parameterisation is taken to mimic the basics of the moderate UV field temperature-extinction profiles of \citet{Adams2004,Facchini2016} and the change of mean-molecular weight from the models of \citet{Storzer1998}. These works show that, in general, the temperature monotonically declines from a roughly constant temperature, where the gas is dominated by atomic hydrogen, to a lower constant temperature where the gas is dominated by molecular hydrogen as the optical depth increases. This transition in temperature and mean molecular weight occurs over column depths to hydrogen of $\sim 10^{21}$--$10^{23}$~cm$^{-2}$.} 

We take the optical depth at the outer boundary of the simulation domain to be zero. For $c_{s,{\rm cold}}$ we assume the disc is molecular, with a mean-molecular weight of 2.35 and has a temperature profile corresponding to the outer regions of protoplanetary discs \citep[e.g.][]{Haworth2016}:
\begin{equation}
T_{\rm cold} = \max\left[100\,{\rm K}\left(\frac{R}{1\,{\rm AU}}\right)^{-1/2},10~K\right]
\end{equation}
whereas for $c_{s,\rm {hot}}$ we assume the gas is dominated by atomic hydrogen with a mean-molecular weight of 1.3 and $T_{\rm hot}=150\,$K. The optical depth is calculated assuming a constant specific opacity $\kappa_{UV}$ to UV photons per unit gas mass. Every time-step in our numerical code we compute the optical depth profile to UV photons using numerical integration of our density profile, this optical depth profile is then used to set the temperature profile (via Equation~\ref{eqn:csprofile}) and like in the isothermal case, we then evolve our simulations to steady-state. Additionally, the inclusion of an optical-depth dependent temperature profile breaks the scale-free nature of our previous isothermal solutions. Thus, we must also specify the gas surface density at the inner boundary. 

Now we have moved to non-isothermal models one of the unclear choices is how to treat the viscosity. If $\alpha$ were a true fundamental constant then the strength of the viscosity should increase linearly with temperature (at fixed position). Thus across our interface from the cold disc and warmer outflow the viscosity would increase markedly, as should it's ability to transport angular momentum. However, there is certainly no evidence to suggest $\alpha$ is a fundamental constant, in fact  simulations indicate the contrary \citep[e.g.][]{Jankovic2021b,Jankovic2021}. Rather the viscosity could vary smoothly, without a temperature induced jump, from the disc to the outflow. Alternatively, if the viscosity arises from the Magneto-Rotational-Instability (MRI), the increase of the gas' ionization state as one moves from the disc to the outflow is likely to increase the value of $\alpha$ \citep[e.g.][]{Bai2011,PerezBecker2011,Mohanty2013,Jankovic2021b,Jankovic2021}, resulting in a very pronounced increase in the viscosity across the disc-wind interface. Interestingly, experiments with changing the viscosity law indicate it has a limited impact on the flow structure, with a more rapidly increasing viscosity with temperature resulting in a marginally wider transition from the disc to outflow. Thus, we choose to adopt the constant alpha model for the remainder of our models, but note this is worth exploring further in future models.

\subsection{Results}

As our starting model we consider an inner boundary at 26.67~AU, with an outer boundary at 600~AU\footnote{Note since we're taking into account the mean-molecular weight change in the non-isothermal models we don't need such a large domain for the same temperatures to include the sonic point.}, we use 1000 grid cells logarithmically spaced between the inner and outer boundaries. At the inner boundary we adopt a surface density of 300~g~cm$^{-2}$, which roughly represents a local disc mass of $\sim 0.02$~M$_\odot$. The surface density is initialised with the pure Keplerian decretion disc profile and then evolved to steady state. 
\begin{figure}
    \centering
    \includegraphics[width=\columnwidth]{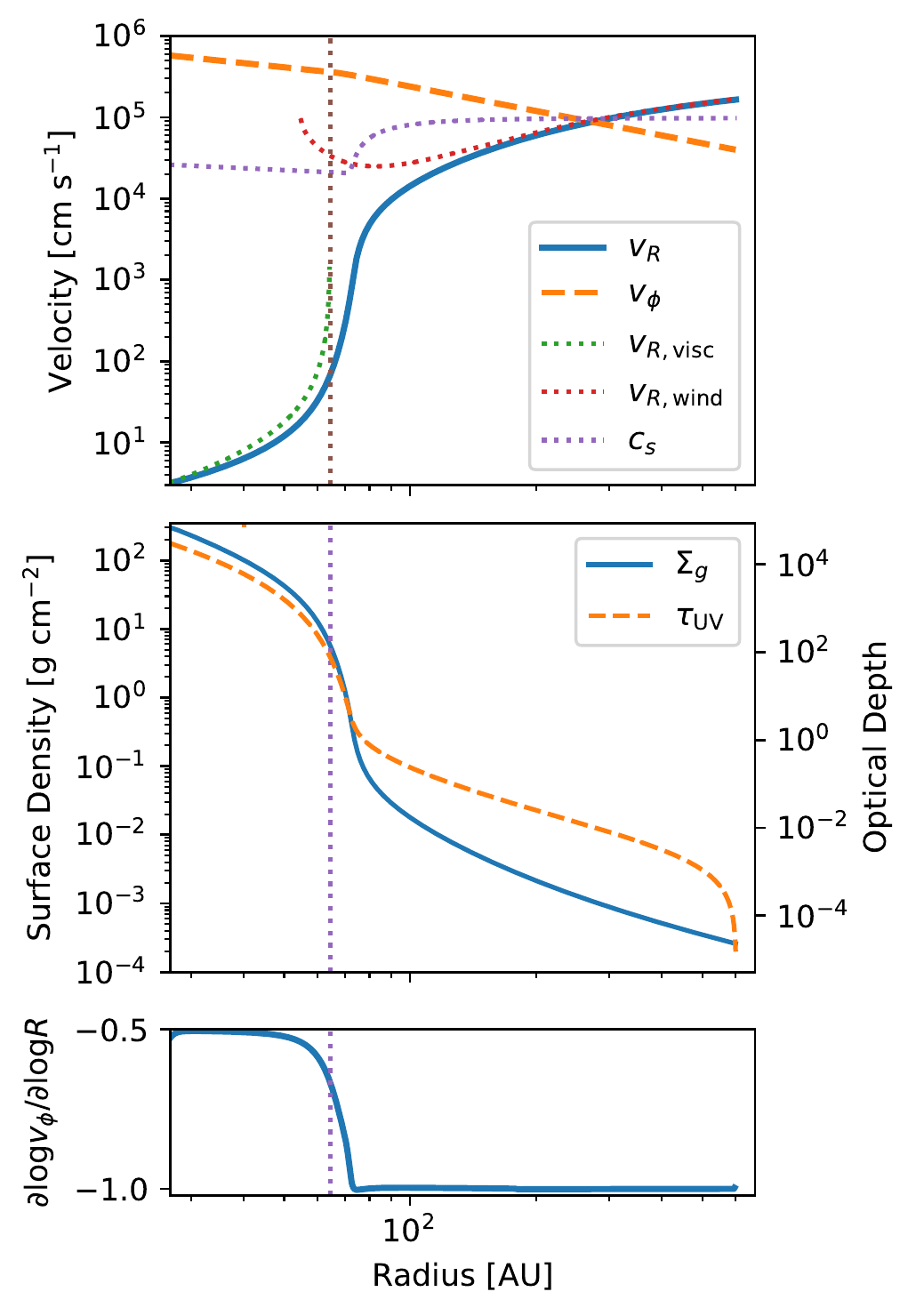}
    \caption{The flow profile for a non-isothermal model with a constant alpha value of $10^{-3}$, $T_{\rm hot}=150~$K and $\kappa_{\rm UV}=15$~cm$^{2}$~g$^{-1}$. The top panel shows the velocity structure, the middle panel shows the density and optical depth and the bottom panel shows the gradient of the azimuthal velocity with radius. he vertical dotted line shows the ``launching'' radius of the wind ($R_L$), defined such that the angular momentum at the outer boundary is $h=\sqrt{GMR_L}$.  }
    \label{fig:noniso1}
\end{figure}
In Figure~\ref{fig:noniso1}, we show the outflow structure for a model using the constant alpha prescription with a value of $10^{-3}$ and $\kappa_{\rm UV}=15$~cm$^{2}$~g$^{-1}$, this value of the opacity roughly corresponds to a UV cross-section of $3\times10^{-23}$~cm$^{2}$ appropriate for an atomic outflow with a dust-to-gas ratio of $\sim 0.01$ that is able to entrain moderately sized dust particles \citep{Facchini2016}. In reality the opacity is set by the actual amount of dust entrained in the outflow (physics we treat in the next section); however, since this is typically dominated by small particles that are readily entrained it is not as coupled a problem as one might imagine. The choice of opacity is rather set by the abundance of small dust grains in the outer regions of the disc. Thus, in this work we choose to treat the opacity as a constant with respect to per unit mass of gas, the dynamical coupling between the disc/outflow, dust entrainment and radiative transfer is an important problem that is worth revisiting within our slim disc framework in future work.  This figure indicates, as in the isothermal model, the disc region is well described by a decretion disc flow profile and the outflow is well described by an isothermal wind model. Interestingly, we see that the ``launching'' radius sits just inside the cold region, before the external heating begins to raise the temperature. While this model has a similar launching radius to the isothermal model shown in Section~\ref{sec:isothermal}, the transition from the Keplerian disc to outflow which conserves specific angular momentum is narrower. This is perhaps unsurprising since the rapid increase in temperature results in a rapid increase in the velocity giving a sharper transition from the disc to outflow. However, the transition is not narrow in terms of flow-scales and still occurs over 4-5 scale heights. 

We now consider perhaps the more pressing question: how does the launching radius change as the value of the viscosity and opacity varies. In previous models that only consider the outflow, the launching radius is exclusively set by the opacity; however, our results from Section~\ref{sec:isothermal} indicate that the strength of the viscosity is likely to play some role. To this end we run calculations with three choices of the viscous alpha: $1\times10^{-4}$, $1\times10^{-3}$ and $1\times10^{-2}$; along with three choices of the opacity: 5, 15 and 45 cm$^{2}$~g$^{-1}$. 

\begin{figure*}
    \centering
    \includegraphics[width=0.99\columnwidth]{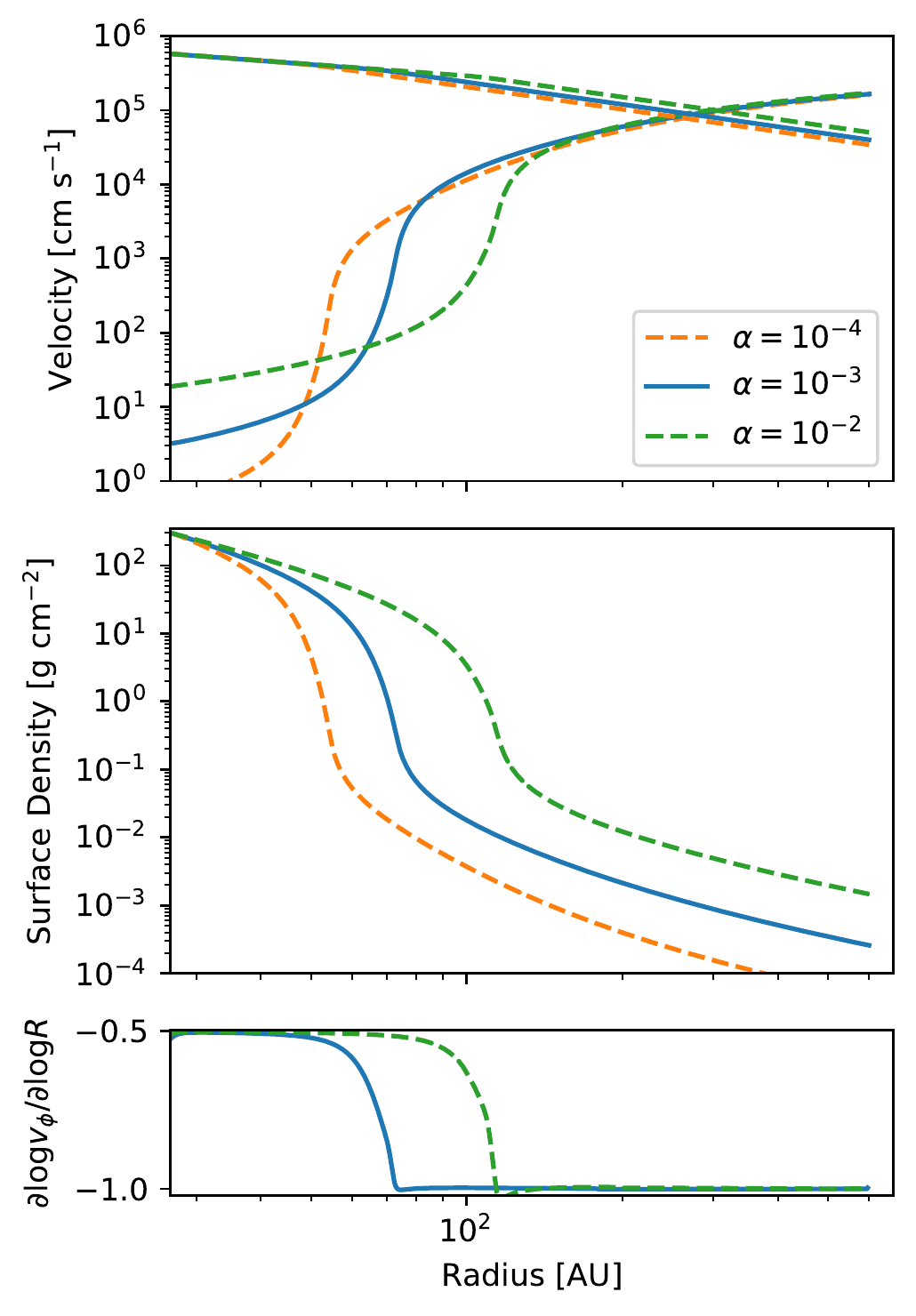}
    \includegraphics[width=0.99\columnwidth]{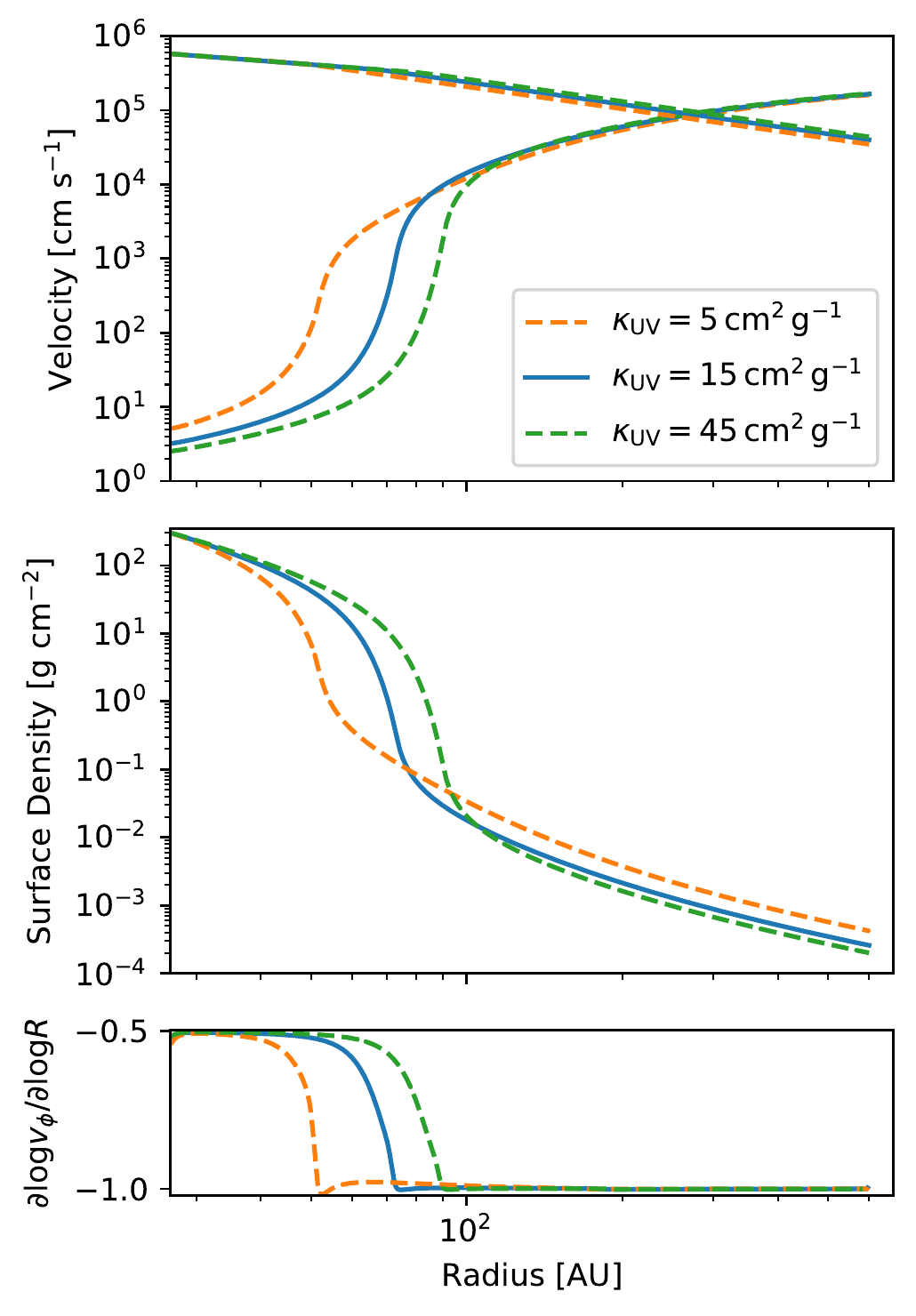}
    \caption{The disc/outflow structure for our non isothermal models where we vary the viscous alpha parameter (left) for a fixed value of $\kappa_{\rm UV}=15$~cm$^2$~g$^{-1}$ and the opacity (right) at a fixed value of $\alpha=10^{-3}$. In all panel the solid blue line is identical to the model shown in Figure~\ref{fig:noniso1}.  }
    \label{fig:non-iso-vary}
\end{figure*}

In Figure~\ref{fig:non-iso-vary}, we show our experiment of varying the viscosity with the opacity fixed to 15 cm$^2$~g$^{-1}$ (left panel) and the results of varying the opacity while fixing $\alpha$ to $10^{-3}$ (right panel). As expected from our isothermal calculations a higher value of the viscosity leads to a larger decretion disc and a larger launching radius as the viscous transport of material is able to supply the outflow with the required mass-loss to large radius. Higher values of the viscosity also yield higher mass-loss rates, however, this naturally arises from our initial conditions which fix the surface density at the inner boundary, since $\dot{M}\propto \nu \Sigma$ in the disc dominated region by fixing the surface density and searching for steady solutions higher values of viscosity naturally yield higher mass-loss rates. In Section~\ref{sec:discuss} we sketch out how to consider our models in the context of an evolving disc.

%

The right-hand panel of Figure~\ref{fig:non-iso-vary} shows that lower values of the opacity allows the outflow to start at higher densities and yields smaller launch radii, again this is as expected as our chosen sound-speed $c_{s,\rm hot}$ is sufficiently high that once the gas reaches this temperature it easily satisfies the $R \gtrsim 0.1 R_{s,0}$ condition found from the isothermal models. Like our initial case in Figure~\ref{fig:noniso1} we find that the transition from the disc dominated region to the outflow dominated region (as indicated in the bottom panels of Figure~\ref{fig:non-iso-vary}) is smooth, but steeper than the isothermal case. Where we typically find the transition is spread over a few scale heights. 

\begin{figure}
    \centering
    \includegraphics[width=\columnwidth]{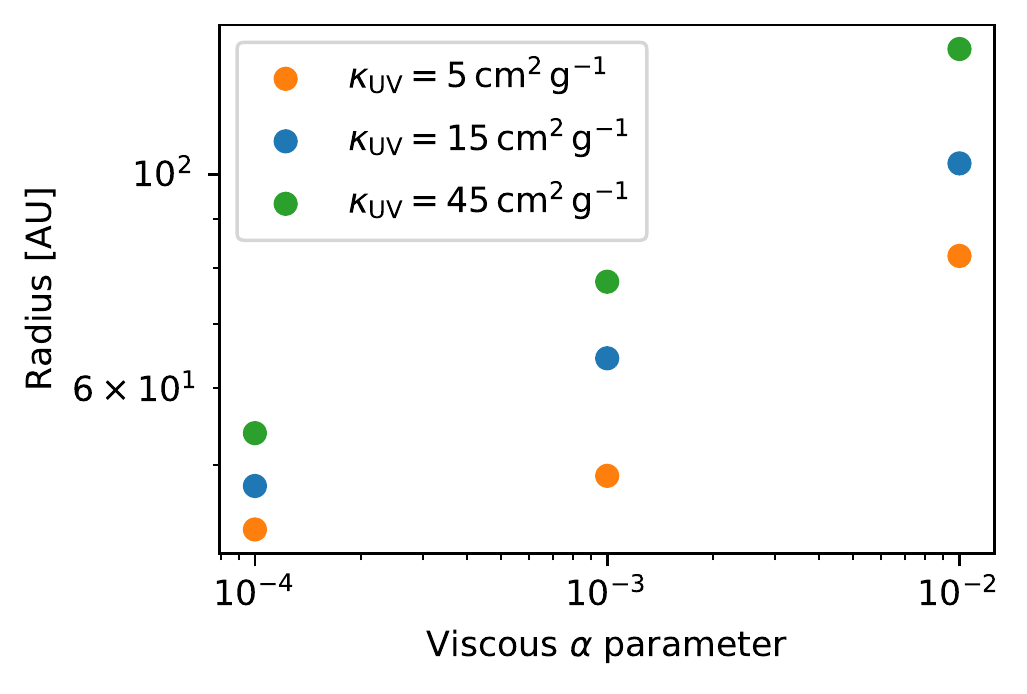}
    \caption{The variation of launch radius, measured from the simulations assuming the angular velocity at the outer boundary yields $\sqrt{GM_*R_L}$, as a function of the viscous alpha parameter for different values of the opacity. Like the isothermal case (shown in Figure~\ref{fig:Rlvsalpha}) we find the launch radius varies logarithmically with viscosity. }
    \label{fig:launch_params}
\end{figure}

While the change in launching radius in our experiment where we vary the opacity and viscosity is similar across all the models we note we have varied the opacity by only a factor of $\sim 10$, where as the strength of the viscosity has changed by $\sim 100$, the variation of the launching radius for the full parameter study is shown in Figure~\ref{fig:launch_params}. This validates our intuition that the properties of the radiative transfer (e.g. opacity, temperature structure) play a dominant role in setting the launch of the outflow, whereas viscosity plays a minor role. Indeed like the isothermal models, the variation of the launch radius with viscosity follows a logarithmic variation with the viscous alpha parameter. 

\section{Dust entrainment in the outflow}

\begin{figure*}
\centering
\includegraphics[width=\textwidth]{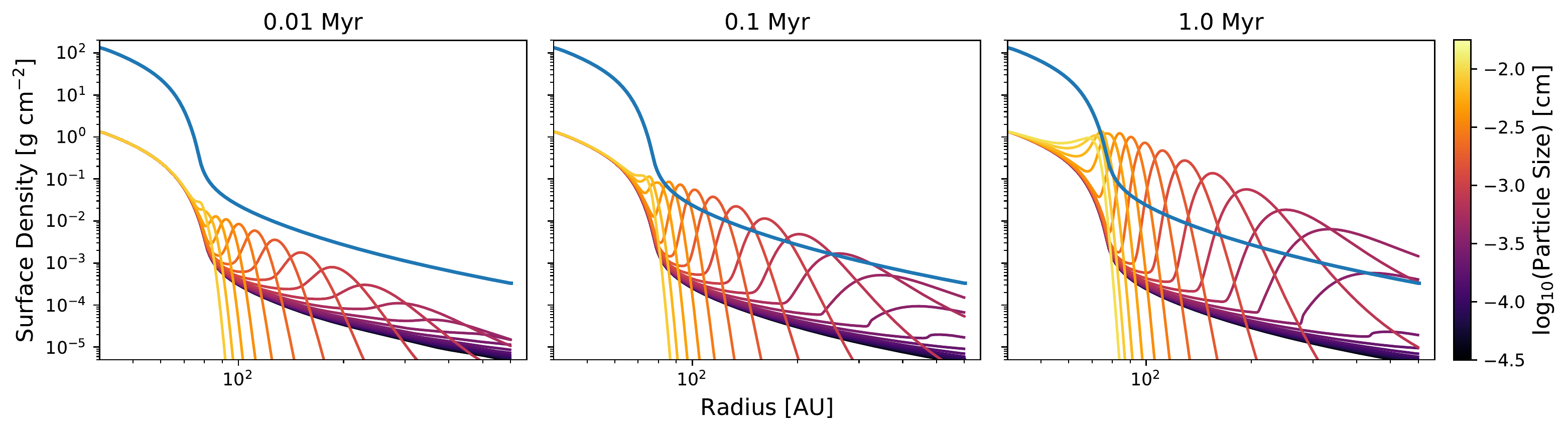}
\caption{The surface density of the gas (thick blue line) and dust particles of different sizes shown as a function of radius and time. The gas is initialised with the steady-state solution and the dust particles are initialised with a surface density of 1\% of the gas.}\label{fig:dustevolve}
\end{figure*}

One of the key questions in the evolution of externally photoevaporating discs is whether dust particles can be entrained in the outflow. In the steady-state limit, when the velocity difference ($\Delta \bm{v} = \bm{v}_{\rm dust} - \bm{v}_{\rm gas}$) between between the gas and dust is small (compared to the gas' velocity) and the dust-to-gas ratio is $\ll 1$, then the velocity difference is well known in the case $v_{R,{\rm gas}}$ is small \citep[e.g.][]{Nakagawa1986,Youdin2005}. For a finite value of $v_{R,\rm gas}$ the velocity differences become:
\begin{eqnarray}
\Delta v_R &= & -  \frac{\eta v_K \tau_s}{1+\xi\left(\frac{v_{R,\rm gas}}{v_k}\right)\tau_s +  \left(\frac{\Omega_{\rm gas}}{\Omega_K}\right)^2\frac{\tau_s^2}{1+\tau_s \left(\frac{v_{R,\rm gas}}{v_k}\right)}}\\
\Delta v_\phi &= & \frac{\eta v_K \tau_s^2\left(\frac{\Omega_{\rm gas}}{\Omega_k}\right)}{2\left[1+\left(\frac{v_{R,{\rm gas}}}{v_k}\right)\left(1+\xi\right)\tau_s+\left\{\left(\frac{\Omega_{\rm gas}}{\Omega_K}\right)^2+\xi\left(\frac{v_{R,{\rm gas}}}{v_k}\right)^2\right\}\tau_s^2\right]}\nonumber
\end{eqnarray}
where $\tau_s$ is the non-dimensional stopping time ($\tau_s = \pi \rho_{\rm in} a / 2 \Sigma_g$ for Epstein drag), $\eta$ is the standard dimensionless strength of the radial pressure gradient $\eta = - 1/(\rho_g\Omega_K^2R)\partial P/\partial R$, \citep[e.g.][]{takeuchi02} and $\xi= \partial \log v_{R,{\rm gas}}/\partial \log R$ and describes how rapidly the radial velocity changes.  For small dust particles ($\tau_s \ll 1$), in the sub-sonic region of the disc/outflow the drift velocities just reduce to the standard results \citep[e.g.][]{jacquet12}:
\begin{eqnarray}
\Delta v_R &= & -  {\eta v_K \tau_s}\\
\Delta v_\phi &= & \frac{\eta v_K \tau_s^2}{2}\nonumber \label{eqn:drift_approx}
\end{eqnarray}
The above result indicates that since $\Delta v_\phi$ is a factor of $\tau_s$ smaller than $\Delta v_R$, implying dust particles will closely follow the angular momentum distribution of the gas. Thus, in-order to determine if the dust is entrained we wish to set $v_{R,{\rm dust}} > 0$ which yields the following inequality for the dust sizes that are entrained in the outflow:
\begin{equation}
    a < \eta^{-1}\frac{\dot{M}}{\pi^2R v_K\rho_{in}} \label{eqn:limit-size}
\end{equation}
This yields a fairly intuitive result that more powerful outflows entrain larger particle sizes, and that steeper pressure gradients will trap smaller grains in the disc (note a similar result was found by \citealt{Facchini2016} and used by \citealt{Selleck2020} in their work). Since $\eta$ is largest in the transition between the disc region and outflow (e.g. Figure~\ref{fig:noniso1}) this is where the limiting dust size is controlled. Therefore, since this is exactly the region where viscosity and the pressure driven outflow are both relevant in setting the surface density and velocity profile the dust entrainment is likely to be more accurately modelled in the slim-disc approach. Specifically, since viscosity tends to smooth out density gradients treatment of the disc/outflow boundary using a continuous model is likely to result in the entrainment of larger dust grains than if a sharp boundary is adopted. 

In addition the viscosity allows the diffusive transport of dust grains outwards, even if $v_{R,{\rm dust}} < 0$, where steep surface density gradients develop allowing turbulent diffusion to transport grains to larger radius. The right-hand side of Equation~\ref{eqn:limit-size} varies with radius as $\propto 1/\eta R^{1/2}$. Since $\eta \sim c_s^2/v_k^2 \partial \log P/ \partial \log R$ which yields  $\propto R\,  \partial \log P/ \partial \log R$ for isothermal flows, we can consider if a dust grain can ever become too big to be entrained at some radius, diffusively transported to another radius, where it becomes entrained again. Such a result is only possible if $\partial \log P/ \partial \log R$ falls faster than $R^{-3/2}$. Inspection of the flow profiles (e.g. Figures~\ref{fig:noniso1} and \ref{fig:non-iso-vary}) indicate that while not definitively ruled out it could only happen over a narrow region of the parameter space where the surface density is changing from rapidly decreasing to falling off as a power-law, just outside the launching radius. Thus, we expect in general diffusive transport will not allow grains that would other wise not be entrained to be transported outwards to regions in the outflow where they can become entrained again. Although, we cannot rule it out entirely for a small range in grain sizes that might stop becoming entrained just outside the launch radius. However, for the simulation parameters we run we do not see such an effect. Additionally, far out in the outflow as it becomes transonic the approximate solution to the drift velocities (Equations~\ref{eqn:drift_approx}) are no longer appropriate, particularly if $\tau_s > 1$ where entrainment becomes very difficult.

\subsection{Numerical method}
To explore this in our numerical calculations we include pressure-less dust fluids in our simulations. The drag due to the gas in both the radial and azimuthal direction is included using the semi-implicit integration method of \citep{Rosotti2016}, thus we do not assume the short-friction time approximation in our numerical calculations. In addition we include a radial velocity due to turbulent diffusion as in \citet{Rosotti2016}, in this work we adopt a Schmidt number of unity for simplicity. With this implementation, advection of the dust's density and momenta can then simply be treated using the same advection scheme used for the gas. The gas densities are low enough and dust sizes we are interested in are small enough that the Epstein drag law is appropriate. We parameterise dust particles in terms of their size, adopting an internal density of 1~g~cm$^{-3}$. 

Finally, we must caution that our treatment of dust as a pressure-less fluid, in particular the diffusive implementation of turbulence, only works if fluid parameters do not change rapidly over a scale-height. In the case of internal EUV photoevaporation where there is a narrow ionization front, the diffusive treatment of turbulence leads to errors in the transport of dust across it \citep[e.g.][]{Booth2021}. However, for our problem of external photoevaporation in intermediate to weak UV regimes we do not get an ionization front (or rarefaction front) where there's a rapid change in fluid properties, but rather a smooth increase in temperature spread over several scale heights. Although, the stopping length of some of the larger particles should be compared to the length scale for fluid properties to change and the inertial size scale of the turbulence.  

\subsection{Results}
Unlike our gas models where we continually evolve the simulations to steady-state we do not evolve the dust distribution to a steady-state. The reason for this is that once the dust drift velocity prevents outwards advection of dust particles, turbulent diffusion can still drive dust outwards. For diffusive transport to compete against the drift requires very steep surface density gradients, which results in unphysically long times to reach steady state and unphysically large dust surface densities. Thus, we take a gas model that has reached steady-state, we then insert dust with a profile such that its initial surface-density is $\Sigma_d = 0.01 \Sigma$. This initial distribution is then evolved forwards for 1~Myr.  

The evolution of the dust surface density as function of time is shown in Figure~\ref{fig:dustevolve} for our standard non-isothermal gas model from Section~\ref{sec:noniso} {\bc ($\alpha=10^{-3}$, $\kappa_{\rm UV}=15$ cm$^2$~g$^{-1}$)}. The panels show that the general picture of entrainment, or not is set early after only $\sim 0.01$~Myr. However, for those dust grains that are not entrained their surface density slowly increases to large values by 1~Myr. For this specific model we find that grains larger than $\sim 50~\mu$m are fully carried out past the sonic point and grains larger than $0.1$mm never actually make it into the outflow as they have an inwards advective velocity even in the disc dominated region. As expected from our discussion above the largest variation in entrained particle size happens at the transition from the disc dominated to outflow dominated region where the radial pressure gradient is largest. This rapid change in entrained particle sizes has observational implications as discussed in Section~\ref{sec:obs_imp}. Since the smallest dust particles are always entrained, and they dominate the UV opacity, then we suspect the selective filtering of the large grains will not impact the radiative transfer, as argued by \citet{Facchini2016}.

\section{Discussion} \label{sec:discuss}

Previous models of externally photoevaporating discs have assumed that the transition from a thin Keplerian accretion disc to a photoevaporative outflow where fluid parcels conserve their specific angular momenta is sharp. In this work, we have used ``slim'' accretion disc models \citep[e.g.][]{Abramowicz1988} to model the actual transition from a disc dominated region to an outflow dominated region. We have focused on intermediate/weakly externally irradiated discs such that the photoevaporation is driven by FUV irradiation and the outflows are warm with a temperature of $\sim$150~K. 

Our results suggest a smooth transition occurring over a several scale heights at the point that incident FUV irradiation begins to be strongly absorbed. We find that the penetration depth of the heating (controlled by the opacity to FUV photons) plays a dominant role in setting the transition from the disc to outflow, and the strength of the viscosity plays a sub-dominant role, only changing the launching radius in a logarithmic manner. In agreement with other works \citep[e.g.][]{Facchini2016}, we find small dust particles can be entrained in the outflow and carried out beyond the sonic point. Larger particles cannot be carried out by the drag force; however, turbulent diffusion can drive particles outwards and if this processes continues for a long $\gtrsim 1$~Myr timescale they can reach large surface densities, given the pressure gradient is steepest in the vicinity of the transition from the disc to outflow, this is where a large range of particle sizes reach the point of not being able to be entrained. 

\subsection{The impact on mass-loss rates}
Accurate mass-loss rates are sensitive to the temperature structure of the outflow, which requires computationally expensive radiative transfer and chemical networks to be solved \citep[e.g.][]{Haworth2018}. Thus, it is important to assess whether the mass-loss rates calculated using {\bc a slim disc approach would differ substantially from those that are calculated using the standard approach.  }

In order to do this, we compute a set of models where we set the viscosity to zero and set the inner boundary of the simulations to the launch radii determined from our slim disc models. At this launch radius we extract the surface densities from our slim disc models and use these as the values in the boundary cells for these zero viscosity calculations. For our standard non-isothermal slim-disc case (e.g. $\alpha=10^{-3}$, $\kappa_{\rm UV}=15$~cm$^{2}$~g$^{-1}$) we find a mass-loss rate of 3.85$\times10^{-8}$~M$_\odot$~yr$^{-1}$ where as for our zero viscosity simulation we find 3.75$\times10^{-8}$~M$_\odot$~yr$^{-1}$. This difference can be understood because the transport of angular momentum in the slim disc model enhances the mass-flux, additionally as the inner regions of the outflow to remain closer to Keplerian which enhances the centrifugal support against gravity, making it slightly easier for the outflow to escape the star's potential increasing the mass-loss rate. We find this trend is replicated across our range of viscosities, where the mass-loss rates for our slim disc models are a few to tens of percent different from those computed using the standard approach. This gives good reassurance that large recomputations of the photoevaporation rates will not be required in the future, and the current models {\bc  \citep[e.g.][]{Haworth2018b} } do provide an accurate representation of actual mass-loss rates. However, if the dust properties are fundamentally changed then recalculation of mass-loss rates maybe necessary. Although, this agreement does not mean our slim disc approach is not necessary; perhaps rather we can connect our slim disc models to current mass-loss models inside the sonic point, {\bc but suitably outside the launch radius, such that viscosity is completely negligible. }

\subsection{Limitations}\label{sec:limitations}

One down-side of our approach is we have only searched for steady-state solutions. For such steady-state solutions to exist there must exist both a source of material at small radius, and a significantly large constant torque to drive the material outwards. Clearly, this is not true in the problem of an externally evaporating protoplanetary disc.  While at large radii, any time-evolving accretion disc does behave like a decretion disc, the expulsion of material and angular momentum outwards is used by the disc to accrete onto the central object.

However, in a real evolving system we expect a situation to arise where as photoevaporation tries to shrink the disc, the disc will try to expand as it accretes. Thus, in reality we expect the disc to quickly evolve to a quasi-steady state where the loss of angular momentum in the outflow is approximately balanced by the loss of angular momentum from the disc and the radius of the disc will slowly evolve \citep[e.g.][]{Anderson2013,Winter2020b}. Since the viscous timescale is typically longest at the outer edge, the exact region we are modelling, we cannot argue that our gas structures represent a situation which will be long-lived over many viscous times. However, we would imagine that the actual structure of an evolving gas disc is not too dissimilar to our gas steady profiles as they reach this steady state on an approximate viscous timescale.

{\bc In this work we have deliberately used a simplified treatment of the radiative transfer, where the sound-speed is parameterised as a function of optical depth. While this treatment picks out the basic features, it is no replacement for real radiative transfer. In particular our approach does not include the fact that the gas temperature is also a weak function of density, as well as a function of extinction. This highlights that further work should aim to incorporate realistic radiative transfer into the slim disc approach.   }

Perhaps, worth discussing in more detail is our dust profiles. As discussed above we do not actually evolve these to steady-state as they will reach unphysical surface densities for dust particles that are too large to be entrained in the outflow. While our simulations show that the dust obtains a general profile that appears to be set over a short timescale, the surface density of the largest particles will continue to increase due to diffusive transport. In particular, we find that the position at which the dust surface density appears to increase is very close to the disc/outflow transition for a large number of dust sizes (due to the steep radial pressure gradient at this position). Thus, if the disc's outer radii is slowly evolving and moving by small distances this could significantly impact the dust surface density distribution in that vicinity. 

In addition, we have included dust diffusion in a manner which simply parameterises turbulence in terms of the dust concentration gradient \citep[following, ][]{clarke88} and assumed it exists in the outflow. It's unclear if this is true, there is limited work as to whether these outflows are turbulent on small scales. Certainly, the multi-dimensional simulations \citep[e.g.][]{Haworth2019} do not show any evidence for turbulence, but they are, by comparison to simulations of studies of hydrodynamic turbulence in accretion discs, low resolution. In addition, these multi-dimensional simulations do not contain magnetic fields. Since ${\rm d}\Omega^2/{\rm d}R < 0$ in the outflows they are in principle unstable to the MRI \citep{balbus91}; however, the MRI can be suppressed if the thermal pressure is low (as would be the case in the outflow, \citealt{Turner2007,Bai2011}). Thus, while we suspect the turbulent transport of dust and the increasing dust-surface profiles associated with the non-entrained grains that are diffusely transported is appropriate for the disc dominated and transition region, it's unclear whether this is realistic for radii well into the outflow.

In addition, we have also not treated dust-growth or fragmentation in any of our calculations. Given the large surface densities of a narrow size range that can be achieved in the transition region this is potentially important as dust grains maybe able to grow, sweeping up smaller particles and preventing their transport into the outflow. \citet{Selleck2020} studied the loss of dust from externally photoevaporating discs adopting the standard method for modelling the disc's evolution and the Birnstiel dust growth model \citep{birnstiel12}. This work found that photoevaporation is able to remove dust mass early in the discs evolution, preventing it from drifting into the inner disc at later stages resulting in a lower dust-to-gas ratio at Myr timescales, with the effect appearing more enhanced for lower mass stars due to their weaker gravitational potentials. The timescale for early loss is set by the timescale for the small grains to grow in the outer regions, thus the potential trapping of larger particles near the disc-outflow boundary that can sweep up small dust particles could reduce the growth timescale and reduce the dust-mass lost. However, more work will need to be done to test this conjecture. 

Finally, in our models we have assumed that the entire model is in hydrostatic equilibrium in the vertical direction. This assumption leads to a super-spherical ($\propto R^{5/2}$) velocity divergence. While this is a good, appropriate approximation in the disc dominated region, the transition region and sub-sonic portions of the flow, it clearly breaks down near the sonic point (where in reality the velocity divergence would become closer to spherical). Thus, by assuming a larger velocity divergence than expected near the sonic point we are slightly underestimating the position of the sonic point and hence overestimating the mass-loss rates. While, this assumption is unlikely to change our flow profile in the regions of interest significantly, the reduction in mass-loss rate may increase the launch radius and reduce the size of the entrained dust particles. In reality, such a problem can only be tackled with multi-dimensional simulations like those presented in \citet{Haworth2019}. 

Thus, while our 1D steady-state models have been informative and elucidated the basic physics, they do not tell us in detail how externally photoevaporating discs evolve and what the dust profile might look like at large radii after millions of years of evolution. Thus, in future work we would advocate solving the time-dependent slim-disc problem for the entire disc structure (perhaps implicitly, allowing large secular time-steps to be taken), or coupling the time-dependent slim-disc problem to a classical thin disc evolution model, but at a radii where the slim-disc model captures the outer region of the disc.

\subsection{Observational Implications}\label{sec:obs_imp}
\begin{figure}
    \centering
    \includegraphics[width=\columnwidth]{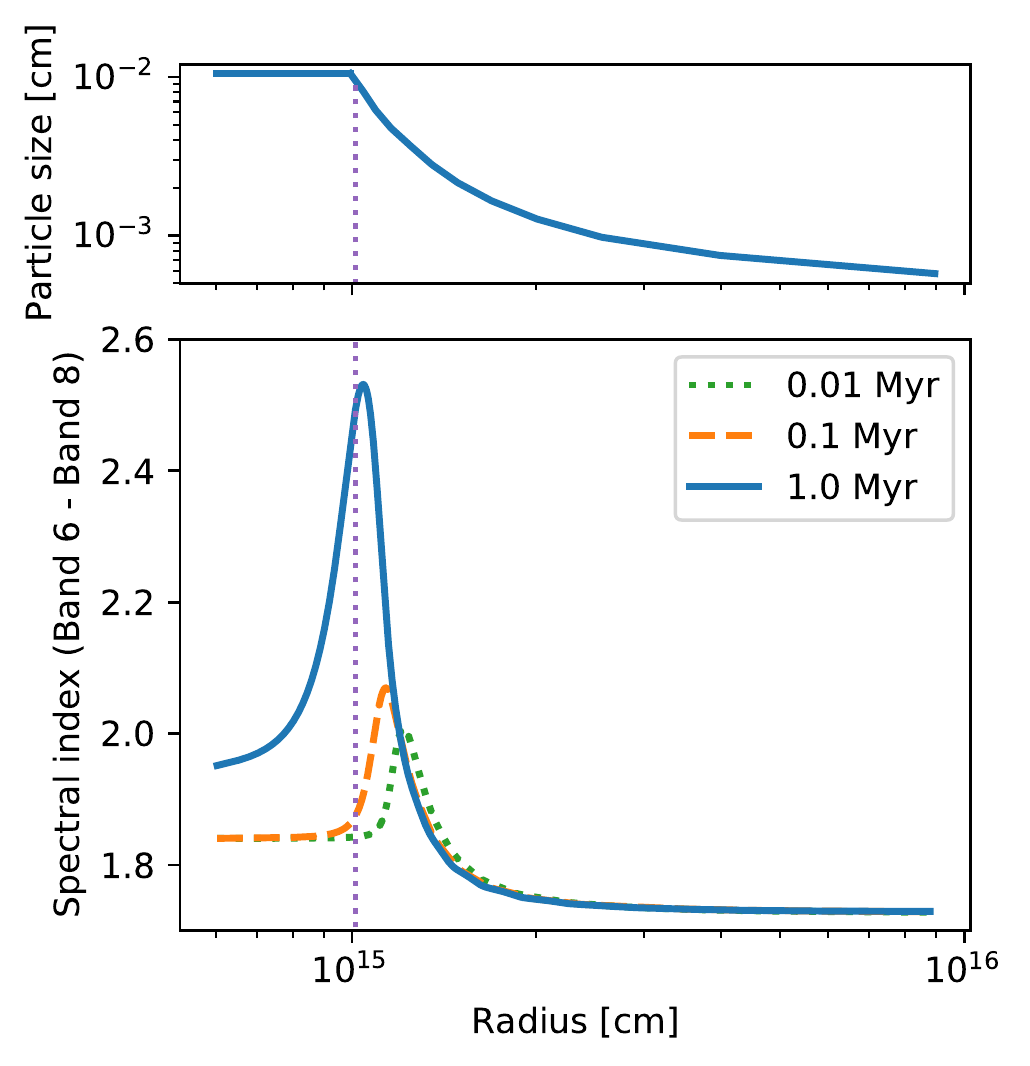}
    \caption{The top panel shown show the maximum particle sizes that has a radial outwards advective velocity as a function of radius and represents the maximum dust particle size that is ``entrained'' for our standard non-isothermal model with dust (Figure~\ref{fig:dustevolve}). The bottom panel shows the spectral index for this model computed between {\it ALMA} bands 6 and 8 as a function of radius and time. The dust particles are assumed to have a minimum size of 0.3~$\mu$m and have an MRN distribution at the inner boundary of the simulations. The launch radius ($R_L$) is shown as the dotted vertical line.}
    \label{fig:spectralindex}
\end{figure}

An interesting question is can we use these models, in combination with state-of-the-art observations to gain insights into the transition between the disc and outflow dominated region. While {\it ALMA} will allow us to study the kinematics of externally photoevaporating discs in detail \citep[e.g.][]{HO2020}, with some recent successes \citep[e.g.][]{Haworth2017,Teague2019}, it's unclear whether the smooth transition found by these models could be confirmed over a model where the angular velocity profile changes abruptly. Although, this might now be possible with sophisticated extraction techniques \citep[e.g.][]{Teague2018a}. Perhaps, more intriguing is how the smooth boundary impacts the dust distribution. In Figure~\ref{fig:spectralindex} we compute the spectral index as a function of radius found if observing our model with {\it ALMA} at bands 6 and 8 (wavelengths of 0.7mm and 1.125mm). These spectral indices are computed assuming the dust is optically thin and that the particle size distribution an the inner boundary of the simulation has an MRN distribution (where the particle size distribution follows $n(a){\rm d}a\propto a^{-3.5}{\rm d}a$. The opacities are computed using the standard ``{\sc dsharp} opacities'' from \citep{Birnstiel2018}. The top panel of this figure shows the maximum entrained dust particle size as a function of radius determined by finding the grain size which has a zero advective radial velocity (i.e. $v_{R,{\rm dust}}=0$). {\bc As expected from our dust surface density plots in Figure~\ref{fig:dustevolve}, the rapid change in the entrained particle size at the transition from the disc dominated region to the outflow gives rise to a change in the spectral index. This change is expected as smaller particles make it into the outflow, while larger particles are retained in the disc. There is also a characteristic bump in the spectral index at the transition radius, where there are large surface densities of a narrow range of intermediate sized particles. In this model this bump is enhanced by the fact this model picks out particles which emphasise this feature in their spectral indices (e.g. 300~$\mu$m, Figure~4, \citealt{Birnstiel2018}). Thus different models will give different spectral index structures, and the net change in spectral index from the disc to outflow will depend on where the spectral index resonance occurs in relation to the particle size distribution. } However, what is clear is that the transition for the disc to outflow should imprint a characteristic feature of a rapidly changing spectral index, along with a change in kinematics. Interesting, evidence for particle size variations have been observed in the externally photoevaporating disc dd114-426 in the Orion Nebula Cluster \citep{Miotello2012} at optical and NIR wavelengths. {\bc While radially decreasing particle sizes are standard outcomes of dust evolution models \citep[e.g.][]{Brauer2008,Birn2010,birnstiel12}, such changes occur smoothly rather than abruptly as in the external photoevaporation case. In addition, the pile up of dust in the transition region should bring the observed dust disc size and gas disc size closer together, rather than further apart as expected in standard dust evolution models in viscous discs \citep[e.g.][]{Trapman2019}. }Thus, including the slim-disc approach in the modelling of synthetic observations for externally evaporating discs could in principle allow detailed analysis and the extraction of parameters of interest such as the magnitude of the viscosity.

\section{Summary}
In this work we have considered the transition from a rotationally supported accretion disc, to a pressure driven outflow where fluid parcels conserve their specific angular momenta. This transition occurs in externally photoevaporating protoplanetary discs, especially in the sub-critical FUV driven case, where there is no ionization front that would cause a sharp jump in temperature and velocity. Using the ``slim'' disc formalism, where advection of linear and angular momentum is explicitly included, we find a smooth transition from the disc dominated region to the outflow dominated region. In the unrealistic situation of isothermal models this transition is found to be quite wide with scales of $\sim R$, similar to the case of outflows driven from Be Star discs \citep[e.g.][]{Krticka2011}. In the more realistic case, where the gas temperature rises into the FUV heated outflow, we find a narrower transition that takes place over several scale heights. We have studied the outflow structure using steady-state discs. As argued in previous works \citep[e.g.][]{Selleck2020}, we find that the penetration depth of FUV photons dominates the launch radius of the outflow; however, we find strength of the viscosity has a minor effect (logarithmic in scaling) on the launch radius. 

In addition by studying the entrainment and diffusive transport of dust we find that the steep density gradient that develops at the transition between the disc and outflow filters out a large range of particle sizes preventing from them being advected into the outflow. However, we find turbulent transport can drive these unentrained particles out to larger radii, leading to high surface densities, which could have important implications for grain growth in the outer regions of the disc. This rapid change in particle size distribution, along with high surface densities leads to a characteristic rapid change in the millimetre spectral index that is potentially observable with {\it ALMA}. 

Our results indicate that accurate modelling the transition from the rotationally supported Keplerian disc to the photoevaporative outflow could have potentially important consequences, particularly for the evolution of the dust and observational implications. Since our work focused on steady-solutions, we motivate further work, where the slim disc formalism is extended to fully time-dependent calculations for the entire disc and outflow.

\section*{Acknowledgements}
We are grateful to the anonymous referee for comments that improved the manuscript. We also thank Thomas Haworth for interesting discussions.
JEO is supported by a Royal Society University Research Fellowship.

\section*{Data Availability}

The evolution code underpinning this article is freely available on GitHub at: \url{https://github.com/jo276/slimdisc.git}.





\bibliographystyle{mnras}
\bibliography{bib_paper}


\bsp	
\label{lastpage}
\end{document}